\documentclass[a4paper]{jpconf}
\usepackage[pdftex]{graphicx}
\usepackage{epstopdf}
\usepackage{amssymb}

\begin{document}
\title{Calorimeter R\&D for the SuperNEMO Double Beta Decay Experiment}
\author{Matthew Kauer - on behalf of the SuperNEMO Collaboration}
\address{Dept. Physics and Astronomy \\University College London \\Gower Street \\London, UK \\WC1E 6BT}
\ead{kauer@hep.ucl.ac.uk}

\begin{abstract}
SuperNEMO is a next-generation double beta decay experiment based on the successful tracking plus calorimetry design approach of the NEMO3 \cite{Arnold:2004xq} experiment currently running in the Laboratoire Souterrain de Modane (LSM). SuperNEMO can study a range of isotopes, the baseline isotopes are $^{82}$Se and possibly $^{150}$Nd. The total isotope mass will be 100--200 kg. A sensitivity to a $0\nu\beta\beta$ half-life greater than $10^{26}$ years can be reached which gives access to Majorana neutrino masses of 50--100 meV. One of the main challenges of the SuperNEMO R\&D is the development of the calorimeter with an unprecedented energy resolution of 4\% FWHM at 3 MeV ($Q_{\beta\beta}$ value of $^{82}$Se). \end{abstract}

\section{Introduction}
\begin{figure}[htp]\centering
\includegraphics[height=8pc]{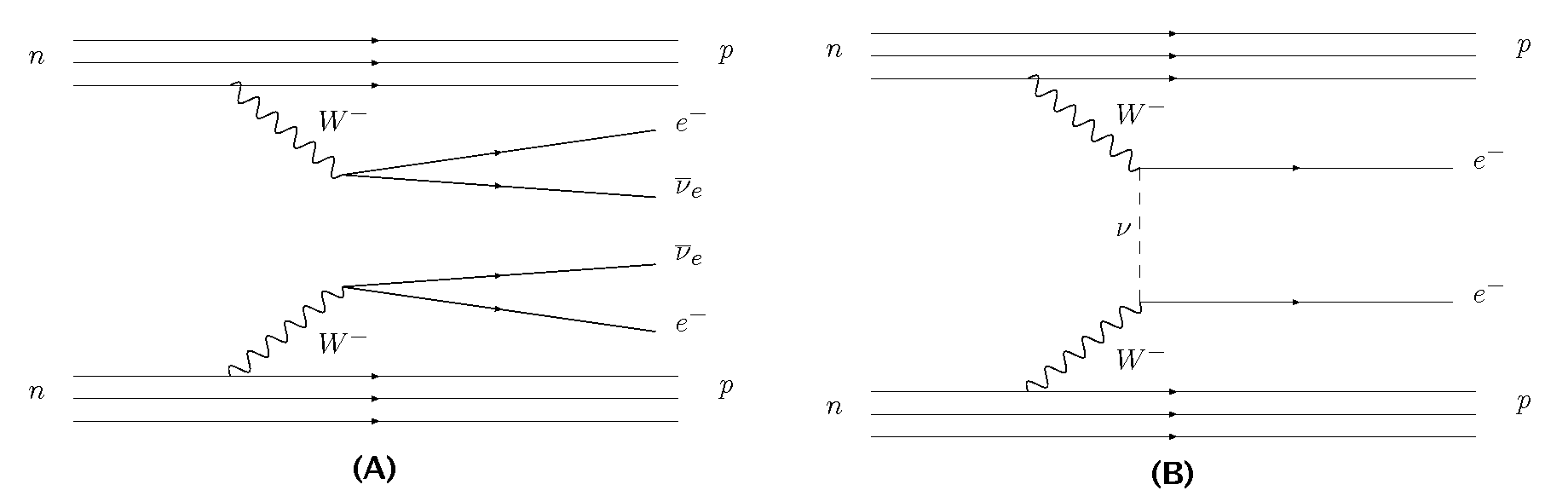}
\caption{Feynman diagrams of a $2\nu\beta\beta$ (A) decay allowed in Standard Model, and $0\nu\beta\beta$ (B) decay allowed if neutrinos are massive and Majorana particles.} \label{fig:betadecays} \end{figure}

The recent observation of neutrino oscillations and the resulting measurements of the neutrino mass differences has motivated experimental searches for the absolute neutrino mass. Neutrinoless double beta decay ($0\nu\beta\beta$) is the only practical way to understand the nature of neutrino mass and one of the most sensitive probes of its absolute value. Ettore Majorana proposed that neutrinos could be their own anti-particles \cite{Majorana:1937vz}, and this lead to Furry's conclusion \cite{Furry:1939qr} that neutrinoless double beta decay (Figure \ref{fig:betadecays}B) is possible via neutrino exchange if the neutrinos are Majorana particles and have non-zero mass. 

The effective Majorana neutrino mass $\langle m_{\beta\beta}\rangle$ is proportional to the square root of the $0\nu\beta\beta$  decay half-life $T_{1/2}^{0\nu}$ in equation (\ref{equ:2b0n}), where $G^{0\nu}$ is the kinematic phase-space factor and $M_{0\nu}$ is the nuclear matrix element. The experimental signature of $0\nu\beta\beta$ is two electrons with the energy sum equaling the $Q_{\beta\beta}$ of the decay. There are other mechanisms to explain neutrinoless double beta decay \cite{Avignone:2007fu}, but the above mechanism is the most favored due to the minimal required modifications to the Standard Model. 

\begin{equation}
[T_{1/2}^{0\nu}]^{-1} = G^{0\nu} \vert M_{0\nu} \vert ^{2} \langle m_{\beta\beta}\rangle^{2} \label{equ:2b0n} \end{equation}

\section{SuperNEMO Detector}
SuperNEMO is $\sim$100~kg source isotope ($^{82}$Se or $^{150}$Nd), tracker + calorimeter detector with a projected neutrinoless double beta decay half-life sensitivity of $10^{26}$ years ($\sim50$~meV effective Majorana neutrino mass). The SuperNEMO baseline design calls for 20 modules ($\sim 4\times 2\times 1$~m), each holding 5 kg of source isotope. Both sides of the source foil have 9 layers of Geiger mode drift cells enclosed by the calorimeter walls. Each module will hold $\sim$600 8" PMTs.

The project is currently in a 3 year design study and R\&D phase and the collaboration comprises over 90 physicists from 12 countries. The R\&D program focuses on four main areas of study: isotope enrichment, tracking detector, calorimeter, and ultra-low background materials production and measurements. The goals of the R\&D are summed up in Table \ref{table:goals}. 

\begin{table}[htp] \centering
\caption{SuperNEMO Parameters and Goals} \label{table:goals} \begin{tabular}{ll} \br
Parameters & Goals \\
\mr
Isotope & $^{82}$Se (or $^{150}$Nd) \\
Mass & 100--200 kg \\
$0\nu\beta\beta$  Detection Efficiency & $30\%$ \\
Energy Resolution FWHM at 3 MeV & $4\%$ \\
$^{214}$Bi Source Purity & $<10~\mu$Bq/kg \\
$^{208}$Tl Source Purity & $<2~\mu$Bq/kg \\
Operation Time & 5 years \\
$T^{0\nu\beta\beta}_{1/2}$ Sensitivity & $10^{26}$ years \\
Effective Majorana Mass $\langle m_{\beta\beta}\rangle$ & 50--100 meV \\
\br \end{tabular} \end{table}

The significance of energy resolution is best illustrated by the half-life sensitivity formula \cite{Avignone:2005cs}. This formula (\ref{equ:sens}) has limitations in accurately predicting the sensitivity of the specific SuperNEMO detector, but does demonstrate the significance of energy resolution. The energy resolution $\Delta E$ factors in with equal importance as isotope mass $M$, runtime $t$,  and number of background events $N_{bkg}$. Factors $N_{A}$ and $A$ are Avogadro's number and atomic mass of the isotope and $\varepsilon$ and $\kappa_{CL}$ are the detector efficiency and the confidence level on the half-life sensitivity $T_{1/2}$. The dominating background to $0\nu\beta\beta$ is the irreducible $2\nu\beta\beta$ channel, therefore the energy resolution of the calorimeter becomes the dominating parameter determining the detector's overall sensitivity to neutrinoless double beta decay. 

\begin{equation}
T_{1/2} \propto \ln 2 \cdot \frac{N_{A}}{A} \cdot \frac{\varepsilon}{\kappa_{CL}} \cdot \sqrt{\frac{M \cdot t}{N_{bkg} \cdot \Delta E}} \label{equ:sens}  \end{equation}

\begin{figure}[htp]\centering
\includegraphics[height=10pc]{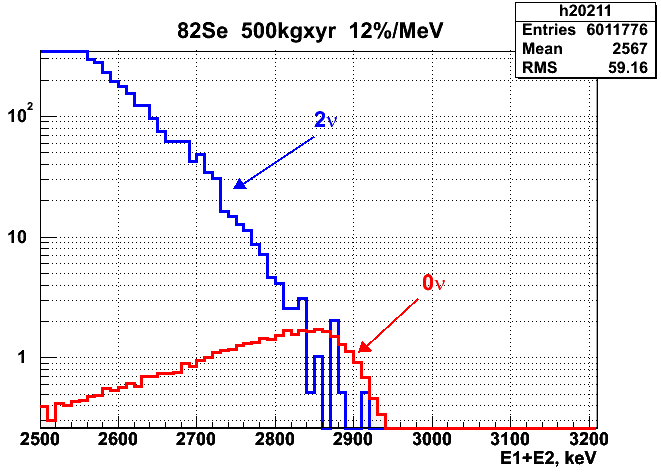}\hspace{3pc}
\includegraphics[height=10pc]{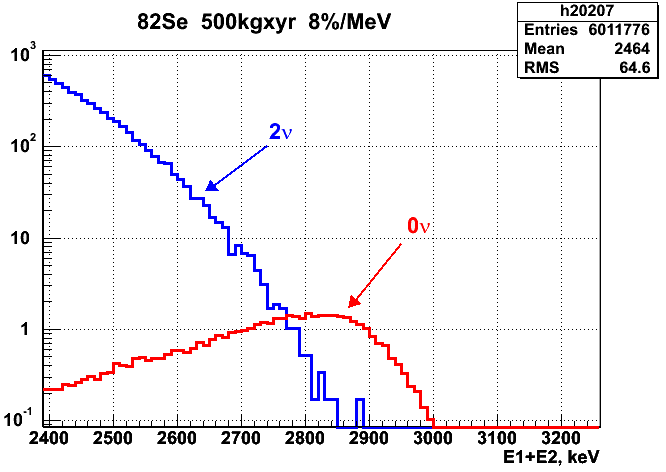}
\caption{Simulations for 500 kg$\cdot$yr $^{82}$Se. The $0\nu\beta\beta$ half-life (RED) is normalized to $10^{26}$ years. Expectations for energy resolutions 12\% (left) and 8\% (right) $\frac{\Delta E}{E}$ FWHM at 1 MeV. } \label{fig:resolution} \end{figure}

Simulations done for $^{82}$Se with a projected calorimeter energy resolution of 12\% and 8\% FWHM at 1 MeV and normalized to $10^{26}$ year $0\nu\beta\beta$ half-life, clearly displays the importance of energy resolution for this experiment (Figure \ref{fig:resolution}). At 12\% energy resolution, the high energy tail from the $2\nu\beta\beta$ energy spectrum overlaps the $0\nu\beta\beta$ peak, but at 8\% energy resolution there is separation. 

\section{Calorimetry Goals of SuperNEMO}
The calorimeter R\&D is subdivided into three main groups: energy and time resolution studies, calibration, and PMT radio-purity. The energy resolution R\&D is the main focus of this report. As with all PMT based calorimeters, PMT gain stability and linearity must be both intrinsically good and experimentally well understood to ensure the accurate reconstruction of data. Conventional LASER/LED configurations prove difficult with many channels. A promising alternative method is one photo-electron peak monitoring \cite{Asch:2005pe} because the PMT gain can be extracted independent of light amplitude. The R\&D also investigates the use a low activity alpha source embedded into the plastic scintillator as a means to monitor the gain.

Specific to low background counting experiments, ultra-pure materials must be used throughout the detector. The PMTs are one of the main sources of contamination with emphasis on the purity of the cathode glass which is closest to the active volume of the detector. The Barium salt used to make conventional glass is chemically the same as Radium, and therefore very difficult to purify during the production of the glass. Various manufactures have developed recipes for low-background glasses, but the requirements of SuperNEMO have motivated this development to a new level of radio-purity. Photonis has provided preliminary samples of their new ultra-pure glass that have met R\&D requirements. 

\subsection{Energy Resolution}
Optimization of the energy resolution is the result of a high number of photo-electrons which reduces the statistical error $1/\sqrt{N_{pe}}$. This can be simplified into three experimental objectives which are described by formula (\ref{equ:photons}). 

\begin{equation}
\frac{N_{ph}}{E_{e}} \cdot \varepsilon_{col}^{light} \cdot \left(QE^{PMT} \cdot \varepsilon_{col}^{PMT}\right)= N_{pe} \label{equ:photons} \end{equation}

$N_{ph}/E_{e}$ is the number of photons per unit energy and is determined by the scintillator light output. $\varepsilon_{col}^{light}$ is the light collection efficiency and depends upon: scintillator geometry, transparency, reflector efficiency, optical coupling quality, etc. Intrinsic characteristics of the PMT include the quantum efficiency of the photo-cathode $QE^{PMT}$, and the cathode to first dynode collection efficiency $\varepsilon_{col}^{PMT}$. There has been a significant breakthrough in development new high quantum efficiency PMTs based on bi-alkali photocathodes by Hamamatsu and Photonis. The SuperNEMO group is working very closely with PMT manufacturers on characterizing these new photo-detectors  which have now a QE in the range of 35--43\% at the peak wavelength (to be compared with $\sim$25\% QE for "conventional" photo-multipliers). Assuming that the energy resolution of the scintillator detector is mainly determined by the photon statistics we can express the resolution in terms of the number of collected photo-electrons (\ref{equ:res1}).

\begin{equation}
\frac{\Delta E}{E}=\frac{FWHM}{E}=\frac{2.35\sigma}{E}=\frac{2.35}{\sqrt{N_{pe}}} \label{equ:res1} \end{equation}

The scintillator must be a low Z material to minimize backscattering electrons and has to have a good timing resolution (a coincidence time resolution of $\sigma$ = 250~ps at 1 MeV is required). It has to be cost effective and radio-pure. These requirements essentially rule out many popular non-organic scintillator, such as NaI(Tl), CsI(Tl), CaF$_{2}$(Eu) etc. which would otherwise provide a good energy resolution due to their high light output. The choice of reflective material is also limited to low density reflectors to reduce electron energy loss through the material.

\section{Experimental Setup}
The energy resolution measurement is carried out by exciting the scintillator under test with a flux of electrons of known energy and then analyzing the resulting distribution. The mono-chromatic source of electrons approximates the delta function and therefore any smearing of the distribution is due to the light collection of the scintillator and PMT under study. The test setup can be broken into three subcategories: the calorimeter block (scintillator + reflector + lightguide + PMT), the electron source, and the data acquisition (DAQ). 

\subsection{Calorimeter Block}
Many different scintillator, reflector, and PMT combinations are being studied. Solid scintillator candidates include polystyrene (PST) based scintillators from ISM and JINR labs (1.5\% PTP, 0.0175\% POPOP) and polyvinyltoluene (PVT) based scintillators from Bicron (BC404, BC408) and Eljen (EJ204, EJ200) manufacturers. Liquid scintillators are toluene based and from CENBG, INR, ISM, and JINR labs (0.5\% PPO, 0.0025\% POPOP). Various specular and diffusive reflectors being tested include: Teflon, Kapton, Aluminized Mylar, and Enhanced Specular Reflector (ESR) from the Vikuiti and ReflechTech manufactures. The three PMT competitors are Hamamatsu, Photonis, and Electron Tubes Ltd. (ETL).

\subsection{Electron Source}
There are two methods used to obtain a mono-chromatic source of electrons. The first method is simplest to implement as one uses the K-shell 976 keV conversion electrons (CE) from a $^{207}$Bi source. The drawback to this method is that the fitting function needs to incorporate the convolution of additional x-rays, gammas, L-shell and M-shell conversion electrons. The second method is more involved to set up, but in principle leaves a spectrum that can be easily fit with a Gaussian function. The $\beta$ emission from a highly active $^{90}$Sr source is passed through a magnetic field so that $\beta$'s of a particular energy can be selected. For the energy resolution measurements, 1 MeV electrons are used. 

\subsection{Data Collecting and Analysis}
Data acquisition is accomplished with a gated QDC (charge to digital converter). The PMT signal is split in two, half the signal is used for triggering of the electronics and generating the gate signal for the QDC, the other half of signal goes directly to the QDC after some passive delay to match the timing of the electronics. In the method of the $^{207}$Bi source, three different data runs must be taken to obtain a pedestal, an energy spectrum of just the gammas (achieved by shielding out the electrons with 2 mm of Aluminum) and the energy spectrum of the gammas + CEs (conversion electrons). The Compton edges from the gamma distribution are sufficiently described by a modified Heaviside step-function. The free parameters of the gamma distribution are determined and then fixed while the gamma + CEs distribution is fit. The CEs are a sum of three Gaussian distributions from the K, L, and M shells.

\section{Measurements}
\begin{figure}[htp] \centering
\includegraphics[height=12pc]{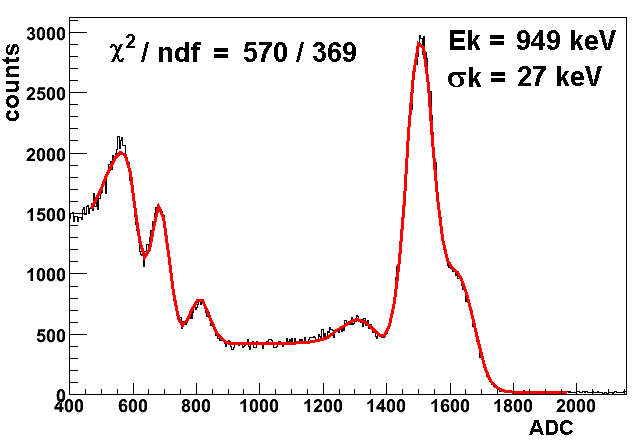}
\caption{The fit to data (RED line) results in 6.5\% FWHM at 1 MeV.} \label{fig:best} \end{figure}

The calorimeter baseline design calls for 8" diameter PMTs, but as a check of physical limitations on achievable energy resolution, a detailed study of small (3") PMTs preceded. A resolution of 6.5\% FWHM at 1 MeV was measured using Bicron BC404 scintillator, wrapped in Vikuiti ESR (Enhanced Specular Reflector), mounted on a 3" Hamamatsu Super-Bialkali type PMT (Figure \ref{fig:best}). Using (\ref{equ:res1}), this extrapolates to 3.8\% at 3~MeV which is better than the goal of 4\% stipulated by the R\&D. This is an unprecedented result for plastic scintillators. Proving there are no physical limitations to reaching the 4\% level, the challenge then becomes scaling up the PMT and scintillator size while maintaining resolution. 

\subsection{Light Collection Simulations with GEANT4}
Extensive optical simulations were carried out in GEANT4 with all inputs being wavelength dependent, experimental measurements including: POPOP absorption and re-emission (Stokes Shifting), PMT QE, scintillator bulk absorption and emission, material indexes of refraction, and material reflectivities. The simulations revealed sensitive parameters of the setup. Polishing the side of the lightguide to give specular internal reflection as well as wrapping the lightguide with a specular reflector yielded a 2--3\% improvement in the expected resolution. The simulations gave expected resolutions of 7.5\% and 7.7\% for the 8" and 11" PMTs with lightguide wrapped in ESR and $5\times 5\times 2$~cm BC404 scintillator wrapped in ESR.

After optimizing the sensitive parameters in our experimental setup, the resolution measurements were 1--2\% worse than the expectations from simulation. Current suspects awaiting investigation are photo-cathode QE uniformity and cathode to first dynode collection efficiency in the presence of Earth's natural magnetic field. Earth's magnetic field is known to influence large ($>$5") PMTs collection efficiency with non-negligible effects, and the effect increases with PMT diameter. These are both characteristics of the PMT which change from one PMT to another and are therefore difficult to simulate accurately.  

\subsection{Solid Scintillator Measurements}
Large solid scintillator blocks are an ideal candidate for SuperNEMO because of low cost, high radio-purity, decreased number of channels, and the physical simplicity of the setup. Three possible variations under study are: small ($<$5") PMT with flat cathode window with scintillator coupled directly to cathode window (Figure \ref{fig:options}A), large ($>$8") PMT with hemispherical cathode window with scintillator coupled to concave lightguide (Figure \ref{fig:options}B), coupled to cathode window, and large ($>$8") PMT with hemispherical cathode window with concave scintillator coupled directly to cathode window (Figure \ref{fig:options}C). Table \ref{table:solids} summarizes the best measurements for these configurations. 

\begin{table}[htp] \centering \caption{Measurements with the Solid Scintillator Setup} \label{table:solids} \begin{tabular}{lll} \br
Scintillator Dimensions & PMT Diameter & FWHM \\
and Type & and Make & at 1 MeV \\
\mr
$5\times 5\times 2$~cm BC404 & 3" Hamamatsu SBA & 6.5\% \\
$9\times 9\times 2$~cm BC408 & 8" Hamamatsu SBA with Lightguide & 10.1\% \\
$14\times 14\times 2$~cm BC404 & 8" Electron Tubes Ltd. with Lightguide & 9.2\% \\
$15\times 15\times 2$~cm BC408 & 8" Hamamatsu SBA with Lightguide & 10.3\% \\
$20\times 2$~cm (hexagonal) BC408 & 8" Hamamatsu SBA with Lightguide & 11.2\% \\
$\varnothing~20\times 2$~cm PST & 8" Photonis & 7.5\% \\
$\varnothing~20\times 10$~cm PST & 8" Photonis & 8.2\% \\
\br \end{tabular} \end{table}

\subsection{Liquid Scintillator Measurements}
\begin{figure}[htp] \centering 
\includegraphics[height=8pc]{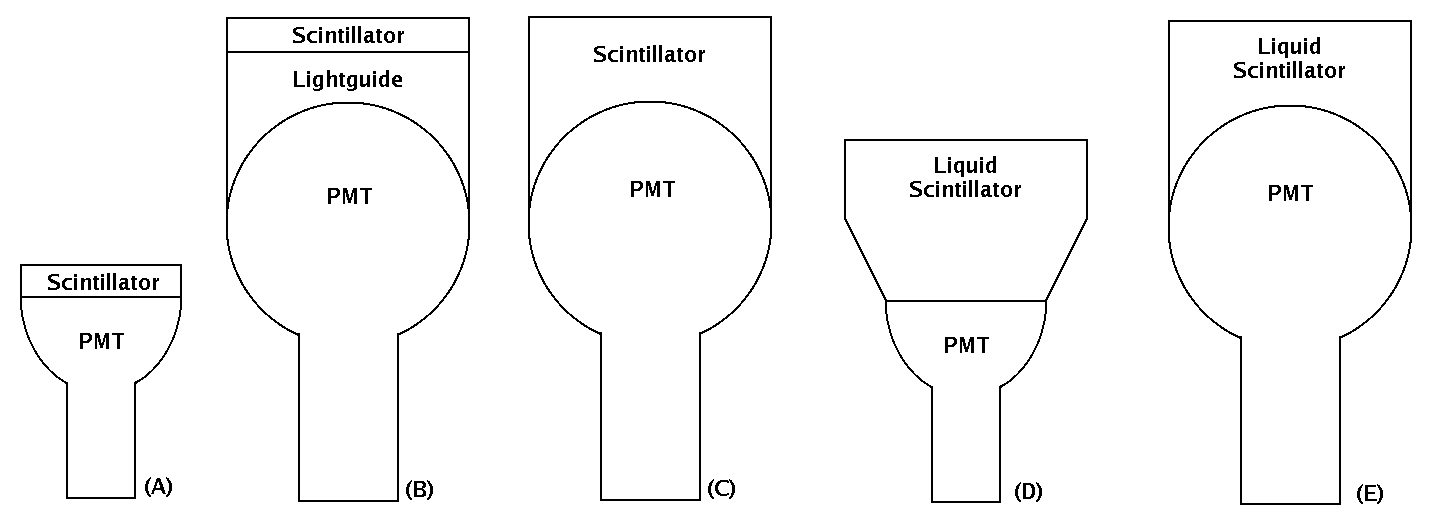}
\caption{Configurations for the solid and liquid scintillator setup.}  \label{fig:options} \end{figure}

In parallel an R\&D program on liquid scintillator detectors is being carried out. The motivations for using liquid scintillator are the following: lower cost, no lightguide needed to couple to hemispherical PMT cathode, larger active volume increases gamma tagging efficiency for background rejection, good uniformity, and high radio-purity. The dominating drawback is the mechanical engineering of the containment structure and meeting safety requirements for an underground laboratory. Two main variations on the setup are under study: a semi-conical setup where the diameter of the liquid scintillator surface is larger than that of the PMT (Figure \ref{fig:options}D), and a cylindrical setup where the diameter of the liquid scintillator surface matches that of the PMT (Figure \ref{fig:options}E). Table \ref{table:liquids} summarizes the best liquid scintillator measurements. 

\begin{table}[htp] \centering \caption{Measurements with the Liquid Scintillator Setup} \label{table:liquids} \begin{tabular}{lll} \br
Scintillator & PMT Diameter & FWHM \\
Dimensions & and Make & at 1 MeV \\
\mr
$\varnothing~7.6\times 2$~cm & 3" Photonis & 7.6\% \\
$\varnothing~7.6\times 10$~cm & 3" Photonis & 8.0\% \\
$\varnothing~8.4\times 9.2$~cm & 5" Photonis & 7.3\% \\
$\varnothing~20.3\times 20$~cm & 8" Photonis & 11.3\% \\
$23\times 9.2$~cm (hexagonal) & 5" Electron Tubes Ltd. & 10.8\% \\
\br \end{tabular} \end{table}

\subsection{Liquid + Solid Hybrid Measurements}
The mechanical engineering of the liquid scintillator containment is challenging. A thin film entrance window with low density and low Z must be used to minimize electron energy losses. An alternative approach is to use a so-called active window where solid scintillator is used on the containment face. This approach utilizes the liquid scintillator as the lightguide and increases the active volume for gamma tagging efficiency. Table \ref{table:hybrids} summarizes the best hybrid measurements. 

\begin{table}[htp] \centering \caption{Measurements with the Liquid + Solid Hybrid Setup} \label{table:hybrids} \begin{tabular}{lll} \br
Liquid / Solid Scintillator & PMT Diameter & FWHM \\
Dimensions & and Make & at 1 MeV \\
\mr
$23\times 9.2$~cm (hexagonal) / $5\times 5\times 2$~cm & 5" Electron Tubes Ltd. & 12.3\% \\
$23\times 9.2$~cm (hexagonal) / $23\times 2$~cm (hexagonal) & 5" Electron Tubes Ltd. & 15.1\% \\
\br \end{tabular} \end{table}

\subsection{Long Bar Scintillator Measurements}
The detector "floor-space" requirement can drastically be reduced by implementing the long scintillator bar design. In this configuration, 2 meter scintillator bars span the tracker volume with a PMT coupled to each end of the bar. This configuration is also the cheapest because of the drastically reduced number of PMTs and the reduced floor-space required from an underground laboratory. With the timing from the two PMTs, an impact resolution of 1--2 cm (along the bar length) is achievable and this information is of additional use for background rejection. Moreover, due to a significantly reduced mass of PMT glass and their relatively remote locations from the detector fiducial volume, the bar design should have a much lower background from PMTs which is one of the main background sources of SuperNEMO. Table \ref{table:bars} summarizes the bar scintillator measurements so far.

A resolution of 7\% at 1 MeV is probably impossible to reach with 2m bars. Thus the crucial question for feasibility of this design is whether a better background rejection and higher detection efficiency compensate a worse energy resolution. Rough estimates show that it might be a valid option if a resolution of 10--11\% is achievable with the bars. Extensive physics simulations are under way to answer this question with certainty. In the meantime measurements are being carried out with high QE PMTs and optimized geometry to reach the best possible resolution with scintillator bars.

\begin{table}[htp] \centering \caption{Measurements with the Bar Scintillator Setup}\label{table:bars} \begin{tabular}{lll} \br
Scintillator Dimensions & PMT Diameter & FWHM \\
and Type and Reflector & and Make & at 1 MeV \\
\mr
$200\times 10\times 1.25$~cm / BC408 / Al. Mylar & 3" Hamamatsu SBA & 12.9\% \\
$200\times 10\times 1.25$~cm / BC408 / Al. Mylar & 3" Hamamatsu SBA with Lightguide & 13.6\% \\
$200\times 10\times 1.25$~cm / BC408 / ESR & 3" Hamamatsu SBA & 12.9\% \\
$200\times 10\times 1.25$~cm / BC408 / ESR & 5" Electron Tubes Ltd. & 13.7\% \\
\br \end{tabular} \end{table}

\section{Summary and Future Plans}
Exceptional resolutions of 6.5\% at 1 MeV were measured for small PVT scintillators coupled to high QE PMTs. The SuperNEMO baseline design calls for large scintillator blocks ($\varnothing~20\times 10$~cm). Scintillators of this size read out through a lightguide showed an energy resolution of 9--10\% at 1 MeV.  Better results have been achieved by casting a large plastic scintillator directly on a hemispherical 8" PMT. With this configuration we have been able to reach the important milestone of 7--8\% $1/\sqrt{E}$ MeV energy resolution for the baseline detector design. Consequently the R\&D on solid scintillators will be focusing on cast scintillator solutions rather than lightguides to increase the light collection efficiency. The development program will also move away from the previous square-block designs and focus on more realistic hexagonal scintillator geometries. We note that there is room for further improvements by using a higher QE PMTs and more efficient scintillators.

Liquid scintillator provides an alternative while maintaining good resolution (7--8\% at 1 MeV) and improving gamma tagging efficiency, but achieving the required resolution with large blocks as well as the engineering of the mechanical design and safety remain a challenge. The hybrid solution creates a more robust containment setup for the liquid, but achieving $<$7\% is very challenging. Long scintillator bars design can potentially give a more efficient detector with more background rejection power. It will drastically reduce the number of PMTs and facilitate a more compact detector design. Measurements so far yield 12--13\% at 1 MeV. Work is in progress to understand if this resolution can be improved to 10\% and whether a worse energy resolution can be compensated by the above advantages of this detector configuration. 

Last years have seen a significant progress in development of novel photo-detectors. PMTs with a QE of over 40\% are now available. Using the latest achievements in PMT, reflector, and scintillator technology the SuperNEMO collaboration has demonstrated the feasibility of achieving the target energy resolution necessary to reach the sensitivity goal of the experiment. The remaining challenge is to demonstrate that the achieved energy resolution can be maintained at the mass production scale. SuperNEMO expects to make the final decision on the calorimeter design in mid-2009. The large scale construction will start in 2011 with the aim to reach the target sensitivity of $\langle m_{\beta\beta}\rangle$~=~50--100~meV by 2017.

\section*{References}
\medskip
\bibliographystyle{unsrt}
\bibliography{./biblio.bib}

\end{document}